# Grapes disease detection using transfer learning

Bhavya Jain[1], Sasikumar Periyasamy[2]

[1]Department of Electronics and Communication Engineering, Galgotias College of Engineering and Technology, Greater Noida, India – bhavyajain516@gmail.com

[2]School of Electronics Engineering, Vellore Institute of Technology, Vellore, India – sasikumar.p@vit.ac.in

**ABSTRACT**

Early and precise diagnosis of diseases in plants can help to develop an early treatment technique. Plant diseases degrade both the quantity and quality of crops, thus posing a threat to food security and resulting in huge economic losses. Traditionally identification is performed manually, which is inaccurate, time-consuming, and expensive. This paper presents a simple and efficient model to detect grapes leaf diseases using transfer learning. A pre-trained deep convolutional neural network is used as a feature extractor and random forest as a classifier. The performance of the model is interpreted in terms of accuracy, precision, recall, and f1 score. Total 1003 images of four different classes are used and 91.66% accuracy is obtained.

**KEYWORDS**

Deep Learning, CNN, Transfer Learning, Image Classification, Random Forest

## 1. INTRODUCTION

Disease identification and classification in plants, in the early stage, is one of the most important agricultural practices. Every year a large number of crops get infected by diseases, resulting in huge economic losses and shortage or increased demand for food in the market. Therefore a quick and robust method has to be made to identify and classify diseases in plants. A lot of advancements have been made in machinery, pesticides used, and other chemicals, but early identification of diseases is the most important aspect. Manually identification is inaccurate, time-consuming, and expensive. In this paper, we have suggested a CNN based model for the early identification and classification of diseases in grapes plants. CNN is the basic Deep Learning tool used in this paper. Deep Learning refers to the use of artificial neural network architectures (ANN), that contain a large number of processing units. The most successful type of models for image analysis to date is convolutional neural networks. CNNs contain many layers that transform their input with convolution filters of a small extent We have used Transfer learning to make a robust and accurate model. We have proposed a CNN based VGG-16 model as a feature extractor and random forest as a classifier. Based on that, we find a confusion matrix and calculated features like precision, recall, and F1 score to evaluate the accuracy of model. These factors can be calculated by using parameters like true positives, true negatives, false positives and false negatives, where true positives and negatives are correctly predicted and false positives and false negatives are wrongly predicted. Precision is the ratio of correctly predicted positive observations to total predicted positive observations. Recall is the ratio of correctly predicted positive observations to all observations in class. F1 score is the weighted average of Precision and

Recall, so it takes both false positives and false negatives into account.

## 2. LITERATURE REVIEW

In [1] authors have used image processing techniques and a Support vector machine classifier to classify diseases in grapes plant. They used 90 images and obtained 89.9% accuracy. In [2] authors proposed an IOT based model using temperature, relative humidity, leaf wetness, and moisture as parameters and predicted grapes diseases using the Hidden Markov model. In [3] authors used transfer learning to classify grape leaf diseases using SSD Mobile Net architecture and R-CNN Inception V2 model. In [4] authors used image processing techniques and an artificial neural network to detect cotton leaf diseases. They used data in four classes and obtained 84% accuracy. In [5] authors used k-means for image segmentation, gray level co-occurrence matrix as feature extractor, and random forest as a classifier. In [6] authors used transfer learning to detect rice plant diseases. They have used AlexNet architecture as feature extractor and support vector machine as image classifier and obtained 91.37% accuracy. In [7] proposed a model based on principal component analysis and backpropagation network. They achieved 94.29% accuracy for two classes grape downy mildew and grape powdery mildew. In [8] diagnosed two grapes diseases using thresholding and anisotropic diffusion to preprocess images and K-means clustering for image segmentation. In [9] authors trained AlexNet and GoogLeNet deep learning models to identify 26 diseases in 14 crop species. They obtained 99.35% accuracy. In [10] authors evaluated CNN performance on two different computer-aided diagnosis applications, namely thoraco-abdominal lymph node detection and interstitial lung disease classification. They also evaluated different under-studied factors of employing deep convolution neural networks to computer-aided detection problems. In [11] color, shape and texture features were extracted from samples of two grape diseases and two wheat diseases. Backpropagation networks, radial basis function, generalized regression networks and probabilistic neural networks were used as the classifiers to identify diseases.

## 3. METHODOLOGY

Deep learning techniques have shown very accurate and precise results in image recognition [12], image segmentation[13], emotion recognition[14], etc. Deep learning technique is an end to end learning, it learns features at different levels of abstraction as layer increases. In transfer learning initialization of CNN weights occur from a pre-trained network. After analysis, it has been found that transfer learning has better performance as compared to training the model from scratch.

In this model, we have used VGG16 and Random Forest. VGG is an acronym for visual geometric group and it is a pre-trained deep convolutional neural network model. It is trained on over 10 million images, learned how to detect generic features from images. VGG16 has 16 layers among them 13 are convolution layers and 3 fully connected layers. The architecture is very simple it has 2 contiguous blocks of 2 convolution layers, followed by max pooling, then it has 3 contiguous blocks of 3 convolution layers, followed by max pooling, and in the last 3 dense layers. Fig[1] shows the architecture of VGG16. But in our model we have used VGG16 as a feature extractor only, the last three dense layers are removed and a Random forest classifier is used. Fig[2] shows our proposed model where VGG16 is used as a feature extractor and Random forest as a classifier. Random forest is a robust and accurate classification technique. It creates a set of decision trees from randomly selected subset of training set. It then aggregates the votes from different decision trees to decide the final class of the test object.

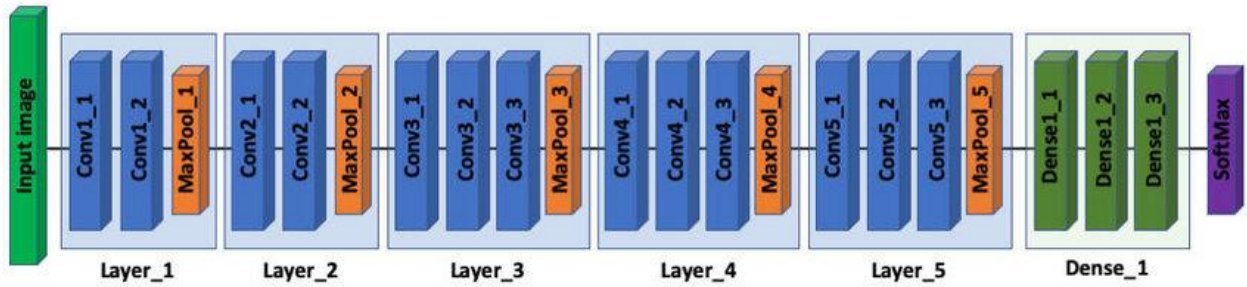

Fig[1] Architecture of VGG16

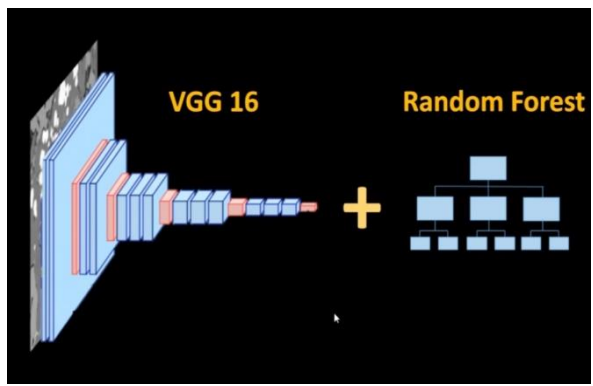

Fig [2] Model Architecture

## 4. RESULT AND DISCUSSION

### 4.1 DATA DESCRIPTION

Total 1003 images from different classes: i) Black Rot; ii) Black Measles; iii) Leaf Blight; iv) Healthy leaves are used. Table 1 shows the description of used data with sample images of each class.

**A) Black Rot:** It is caused by the fungus Guignardia bidwellii. Black rot occurs in hot and humid weather. It attacks all the green parts – leaves, shoots, and fruit.

**B) Black Measles:** It causes superficial spots and on the fruit and leaf. Black Measles or Spanish Measles has long plagued grape growers and its cryptic expression of symptoms. During the season, the spots may coalesce over the skin surface, making the fruit black in appearance.

**C) Leaf Blight:** It is caused by the fungus Helminthosporium turcicum pass. Leaf blightgenerally occurs in humid climates by producing reddish-purple or tan spots on the leaves. It attacks seedlings and older plants.

| Class Name | Number of Images | Image Sample |
|---|---|---|
| Black Rot | 297 | |
| Black Measles | 312 | |
| Leaf Blight | 303 | |
| Healthy Leaves | 300 | |

Table 1. Details of four classes with sample images of each class

### 4.2 EXPERIMENTAL SETUP

The proposed model is implemented in the spyder python environment, using VGG16 as a feature extractor and random forest as a classifier. Total 1003 images of grapes plants are used. The dataset is divided into four classes Black Rot, Black Measles, Leaf Blight, and Healthy leaves. In the image acquisition phase images are resized and converted into RGB format. Then the list is converted into an array and normalized pixel values between 0

to1. Then VGG16 is used for extracting the features, without changing the weights and the random forest is used for image classification.

The performance of the system is interpreted in terms of accuracy, precision, recall, and f1 score. The accuracy for the 80% - 20% training – testing dataset is 91.66%. In Fig [3] Confusion matrix is shown.

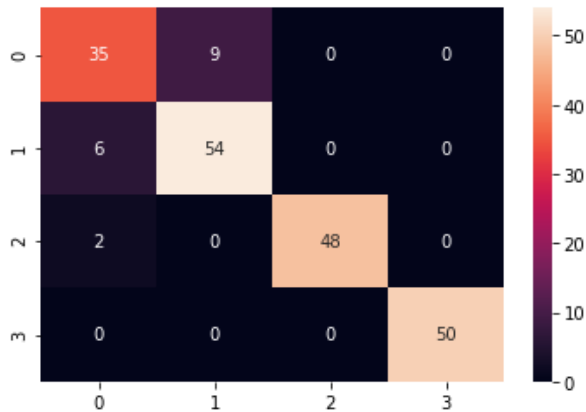

Fig [3] Confusion Matrix

Table 2. shows precision, recall, f1 score of all four classes in 80% - 20% training – testing dataset for all four classes.

| Class | Precision | recall | f1 score |
|---|---|---|---|
| Black Rot | 0.81 | 0.80 | 0.80 |
| Black Measles | 0.86 | 0.90 | 0.88 |
| Leaf Blight | 1.00 | 0.96 | 0.98 |
| Healthy Leaves | 1.00 | 1.00 | 1.00 |

Table 2. Precision, Recall and f1 scores of four classes for 80%-20% dataset

Accuracy of the model is also calculated for different training - testing divisions and shown in Table 3.

| Training-Testing Division (%) | Classification Accuracy (%) |
|---|---|
| 60-40 | 89.21 |
| 70-30 | 90.68 |
| 80-20 | 91.66 |

Table 3. Classification accuracy for different training-testing ratios.

## 5. CONCLUSION AND FUTURE SCOPE

Limited work have been done in quick and automatic disease identification and classification in plants. In this paper transfer learning based model has been proposed to classify diseases in grapes plants. The model has obtained 91.66% accuracy in 80% - 20% training – testing dataset. Moreover, a confusion matrix is produced to evaluate the model. is The performance of the model can be further improved using large datasets.